\documentclass[aip,apl,twocolumn,superscriptaddress,reprint]{revtex4-2}
\usepackage{amsmath} 
\usepackage{bm}  
\usepackage{graphicx}
\usepackage{dcolumn}
\usepackage{physics}
\usepackage{color}
\usepackage[colorlinks=true, citecolor=blue, linkcolor=blue, urlcolor=blue]{hyperref}
\usepackage[T1]{fontenc}
\usepackage{tgtermes}

\graphicspath{ {./}{./figs/} }

\newcommand{\cvs}{\text{cm}^2/ \text{V}\,\text{s}}

\begin{document}

\title{Free-carrier screening unlocks high electron mobility in ultrawide bandgap semiconductor CaSnO$_3$}

\author{Jiayi Gong}
\affiliation{College of Physics, Chengdu University of Technology, Chengdu 610059, China}
\affiliation{Centre for Quantum Physics, Key Laboratory of Advanced Optoelectronic Quantum Architecture and Measurement (MOE), School of Physics, Beijing Institute of Technology, Beijing 100081, China.}

\author{Chuanyu Zhang}
\email{zhangchuanyu10@cdut.edu.cn}
\affiliation{College of Physics, Chengdu University of Technology, Chengdu 610059, China}

\author{Wenjie Hu}
\affiliation{Centre for Quantum Physics, Key Laboratory of Advanced Optoelectronic Quantum Architecture and Measurement (MOE), School of Physics, Beijing Institute of Technology, Beijing 100081, China.}

\author{Jin-Jian Zhou}
\email{jjzhou@bit.edu.cn}
\affiliation{Centre for Quantum Physics, Key Laboratory of Advanced Optoelectronic Quantum Architecture and Measurement (MOE), School of Physics, Beijing Institute of Technology, Beijing 100081, China.}

\date{\today}

\begin{abstract} 
Alkaline earth stannates have emerged as promising transparent conducting oxides due to their wide band gaps and high room-temperature electron mobilities. Among them, CaSnO$_3$ possesses the widest band gap, yet reported mobilities vary widely and are highly sample-dependent, leaving its intrinsic limit unclear.
Here, we present \textit{ab initio} calculations of electron mobility in CaSnO$_3$ across a range of temperatures and doping levels, using state-of-the-art methods that explicitly account for free-carrier screening in electron-phonon interactions. 
We identify the dominant limiting mechanism to be the long-range longitudinal optical phonon scattering, which is significantly suppressed at high doping due to free-carrier screening, leading to enhanced phonon-limited mobility.
While ionized impurity scattering emerges as a competing mechanism at carrier concentrations up to $\sim$~10$^{20}$~cm$^{-3}$, the phonon scattering reduction dominates, yielding a net mobility increase with predicted room-temperature values reaching about twice the highest experimental report. 
Our work highlights the substantial untapped conductivity in CaSnO$_3$, establishing it as a compelling ultrawide bandgap semiconductor for transparent and high-power electronic applications. 
\end{abstract}

\maketitle

\label{sec:intro}
Ultrawide bandgap semiconductors, with energy bandgaps exceeding 4~eV~\cite{Higashiwaki2021_apl}, are critical for applications in deep-ultraviolet (DUV) optoelectronics, high-power electronics, and transparent conducting devices~\cite{tsao2018ultrawide,Yang2022,Muramoto_2014,Varley2022}.
Identifying materials that combine a wide band gap with high conductivity remains a central challenge~\cite{Xu2022}.
Among the promising candidates, alkaline-earth stannates, XSnO$_3$ (X = Ba, Sr, Ca), have attracted attention due to their favorable band gaps and high electron mobilities~\cite{Lee2017,Prakash2019}. 
These materials exhibit a unique electronic structure with the conduction band minimum primarily derived from Sn 5$s$ orbitals, resulting in low electron effective masses and superior conductivity compared to conventional $d$-band perovskite oxides~\cite{zhang2007structural,Stanislavchuk2012}.
Within this family, CaSnO$_3$ stands out with the widest band gap ($\sim$4.6~eV)\cite{liu2023doping}, considerably larger than that of BaSnO$_3$ ($\sim$3.1~eV)\cite{prakash2017wide} and SrSnO$_3$ ($\sim$4.1~eV)\cite{truttmann_combined_2021,liu2024deep}, along with enhanced structural stability in its orthorhombic phase~\cite{Redfern_2011}. 
These features make it a compelling candidate for high-power electronic and DUV applications. 
However, experimentally reported electron mobilities for CaSnO$_3$ vary dramatically across studies---from below $4~\cvs$~\cite{shaili2021higher} to $40~\cvs$~\cite{liu2023doping} ---with this variation highly sensitive to sample quality and doping effectiveness, leaving the intrinsic mobility limit unresolved.\\
\indent
Accurate first-principles calculations of phonon-limited carrier transport provide a powerful means to predict the intrinsic mobility limits of semiconductors~\cite{claes_phonon-limited_2025}.
In recent years, substantial progress has been made in modeling electron-phonon (e-ph) interactions across a wide range of materials, including semiconductors with polar and quadrupolar interactions\cite{,VerdiG,SjakstePRBdfp,BruninGaN,ZhouPRL-GaN}, complex oxides with strong lattice anharmonicity\cite{ZhouSTO}, and low-dimensional systems\cite{Sohier2016}. 
Combined with the Boltzmann transport equation framework, these methods have enabled detailed studies of intrinsic transport properties in materials such as GaAs~\cite{ZhouGaAs,liu2017prb,wuli2018prb}, GaN~\cite{ZhouPRL-GaN,PonceGaN2019}, GeO$_2$~\cite{Bushick2020}, and organic semiconductors\cite{lee2018prb}, among others.
However, most studies focus on low-doping regimes where free-carrier screening effects are negligible and typically neglected. While a recently proposed first-principles approach enable the inclusion of free-carrier screening in e-ph interactions~\cite{macheda_electron-phonon_2022}, its application to doping-dependent transport calculations remains largely unexplored.\\
\indent
In this work, we present comprehensive \textit{ab initio} calculations of electron mobility in doped CaSnO$_3$. 
We implement and extend state-of-the-art methods to explicitly include free-carrier screening in e-ph interactions and charge transport calculations, enabling accurate prediction of phonon-limited mobility as a function of carrier concentration.
Our results show that mobility is primarily limited by longitudinal optical (LO) phonon scattering, driven by strong long-range e-ph coupling. 
At high doping levels, this scattering is significantly suppressed due to free-carrier screening, leading to enhanced phonon-limited mobility. Meanwhile, 
ionized impurity scattering becomes increasingly important at elevated carrier concentrations, contributing to mobility degradation. 
By accounting for both mechanisms, we find that the room-temperature electron mobility can reach nearly twice the highest experimentally reported values. 
Our work establishes the intrinsic mobility limits of doped CaSnO$_3$ while providing microscopic insights into dominant scattering processes, offering theoretical guidance for optimizing conductivity in ultrawide bandgap semiconductors.

\begin{figure*}[!htbp]
\centering
\includegraphics[width=2.0\columnwidth]{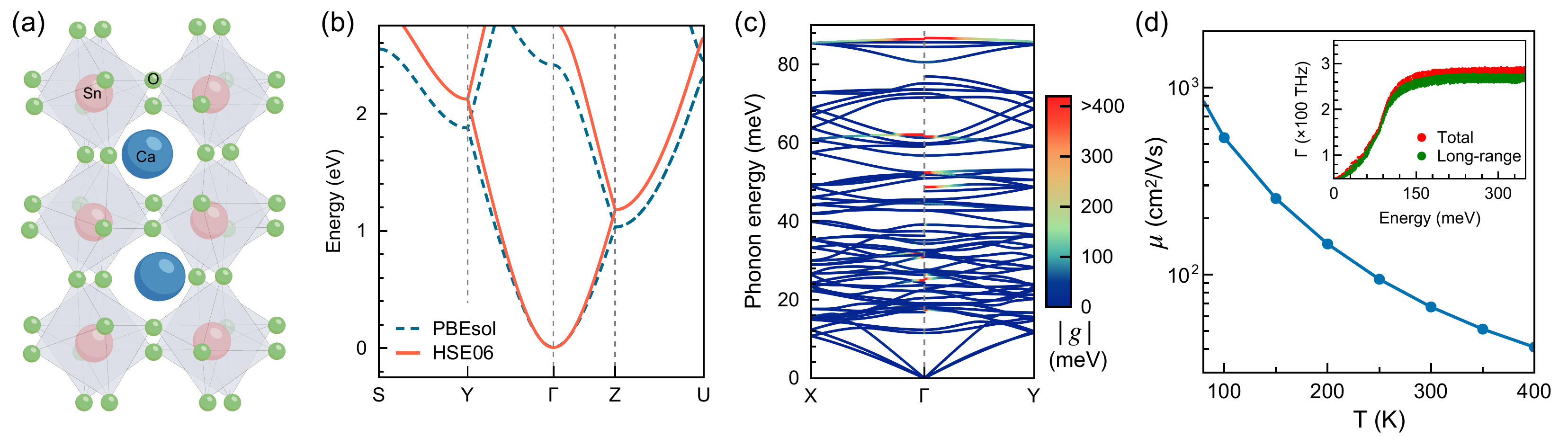}
\caption{(a) Crystal structure and (b) electronic band structure of orthorhombic CaSnO$_3$. The energy of the conduction band minimum (CBM) locate at $\Gamma$-point is set to zero. (c) Phonon dispersion with a color map indicating e-ph coupling strength $|g_\nu(\bm{q})|$. (d) Calculated phonon-limited electron mobility at low doping ($n_c = 10^{17} \text{cm}^{-3}$) as a function of temperature. The inset shows the corresponding scattering rate at room temperature, with the long-range contribution highlighted.}\label{Fig:fig1}
\end{figure*}

\label{sec:methods}
CaSnO$_3$ crystallizes in an orthorhombic structure consisting of corner-sharing SnO$_6$ octahedra, with Ca$^{2+}$ ions occupying the interstitial sites, as illustrated in Fig.~\ref{Fig:fig1}(a). 
The octahedra exhibit characteristic tilts and rotations, leading to a distorted perovskite framework that directly influences the electronic and vibrational properties. 
We carry out density function theory (DFT) calculations on orthorhombic CaSnO$_3$ using the \textsc{Quantum espresso} package\cite{Giannozzi_2009}. 
The fully relaxed lattice constants are $a$ = $5.508$~\AA, $b$ = $5.693$~\AA, and $c$ = $7.956$~\AA, which show excellent agreement with the experimental results ($a$ = $5.514$~\AA, $b$ = $5.663$~\AA, and $c$ = $7.882$~\AA)~\cite{zhaoCaSnO32004}. We employ norm-conserving pseudopotentials from PseudoDojo~\cite{Dojo}, the PBEsol generalized gradient approximation (GGA) for the exchange-correlation functional~\cite{PBE}, and a plane-wave kinetic energy cutoff of 85~Ry to obtain the ground-state electronic structure. 
Lattice dynamical properties and e-ph matrix elements are computed using density functional perturbation theory (DFPT) on coarse Brillouin zone grids of $8\times 8\times 6$ $\bm{k}$-points and  $4\times 4\times 3$ $\bm{q}$-points. These quantities are first transformed to real space in localized basis and then interpolated onto dense $\bm{k}$- and $\bm{q}$-point meshes required for converged transport calculations~\cite{Gonze1997,Giustino2007}. For CaSnO$_3$, we solve the Boltzmann transport equation iteratively on dense $80\times 80\times 80$ $\bm{k}$- and $\bm{q}$-point grids to obtain the mobility as a function of temperature and doping concentration using the \textsc{Perturbo} package~\cite{Perturbo}. The convergence tests are provided in the supplementary material. \\
\indent
In polar materials like CaSnO$_3$, both the dynamical matrix and e-ph matrix elements are decomposed into short-range and long-range contributions that require distinct computational treatments. 
The short-range part of the e-ph matrix element, $g^S$, is obtained through standard Wannier interpoloation. 
The long-range part, $g^L$, arising from macroscopic electric fields generated by LO phonons, is typically computed in the intrinsic limit as\cite{VerdiG,SjakstePRBdfp}:
\begin{equation} \label{Eq:eq1}
 g_{\kappa\alpha}^{L}(\bm{k},\bm{q})   =    \frac{ie^{2}}{\Omega} \sum_{\bm{G}\neq \bm{-q}} \frac{\left[ (\bm{q}+\bm{G}) \cdot \bm{Z}_\kappa^{*}\right]_{\alpha} e^{-i(\bm{q}+\bm{G})\cdot\bm{\tau}_{\kappa}}}{(\bm{q}+\bm{G})\cdot \bm{\epsilon_{\infty}} \cdot(\bm{q}+\bm{G})}  
\end{equation} 
where $\Omega$ is the unit cell volume,  $\bm{G}$ denotes the receprocal lattice vector, $\kappa$ indexes atoms, and $\alpha$ represents Cartesian directions. $\bm{Z}_\kappa^{*}$ and $\bm{\epsilon_{\infty}}$ are the Born effective charge tensor and high-frequency dielectric tensor, respectively. 
These matrix elements are evaluated in the Wannier gauge under the smooth phase approximation following the formalism implemented in \textsc{Perturbo}~\cite{Perturbo}, with electronic band indices omitted for clarity. While this expression accurately describes undoped or lightly doped systems where free-carrier screening is negligible, it becomes inadequate at higher carrier concentrations. \\
\indent
In moderately and heavily doped semiconductors, free carriers substantially modify the dielectric response, screening the macroscopic electric fields that mediate long-range part of the dynamical matrix and e-ph coupling. To capture these effects, a recent theoretical development~\cite{macheda_electron-phonon_2022} extended Eq.~\eqref{Eq:eq1} by replacing the static high-frequency dielectric tensor with a doping- and temperature-dependent dielectric function, $\epsilon^{-1}(\bm{q}, n_c, T)$. Within static linear-response theory, this dielectric function is computed from first principles as: 
\begin{align}
  \delta \chi^{0}(\bm{q}, n_c, T) & = \frac{2}{\Omega} \sum_{m n \bm{k}} \frac{\delta f_{n \bm{k}}-\delta f_{m \bm{k}+\bm{q}}}{\varepsilon_{n \bm{k}}-\varepsilon_{m\bm{k}+\bm{q}}}\left|\langle u_{n \bm{k}} | u_{m \bm{k}+\bm{q}}\rangle\right|^{2}, \nonumber \\
\epsilon^{-1}(\bm{q}, n_c, T) & = \frac{1}{\epsilon_{\infty}-4 \pi e^{2} / q^{2} \delta \chi^{0}(\bm{q}, n_c, T)} \label{Eq:eq2}
\end{align}
where $\delta f $ represents the change in electronic occupation upon introducing free carriers at concentration $n_c$, and $\delta \chi^{0}$ denotes the corresponding change to the independent-particle polarizability. $\varepsilon_{n \bm{k}}$ and $u_{n \bm{k}}$ are the electron energies and periodic part 
of the Bloch wavefunctions, respectively. 
We have developed an efficient computational scheme within \textsc{Perturbo}~\cite{Perturbo} that evaluates ($n_c$, $T$)-dependent dynamical matrix and e-ph matrix elements on ultrafine $\bm{k}$- and $\bm{q}$-point grids at tractable computational costs, and integrates them into transport calculations. 
This implementation enables accurate first-principles calculations of phonon-limited transport properties in doped materials. A detailed explanation of the transport calculations including free-carrier screening can be found in the supplementary material.
\begin{figure}[!htbp]
\centering
\includegraphics[width=\columnwidth]{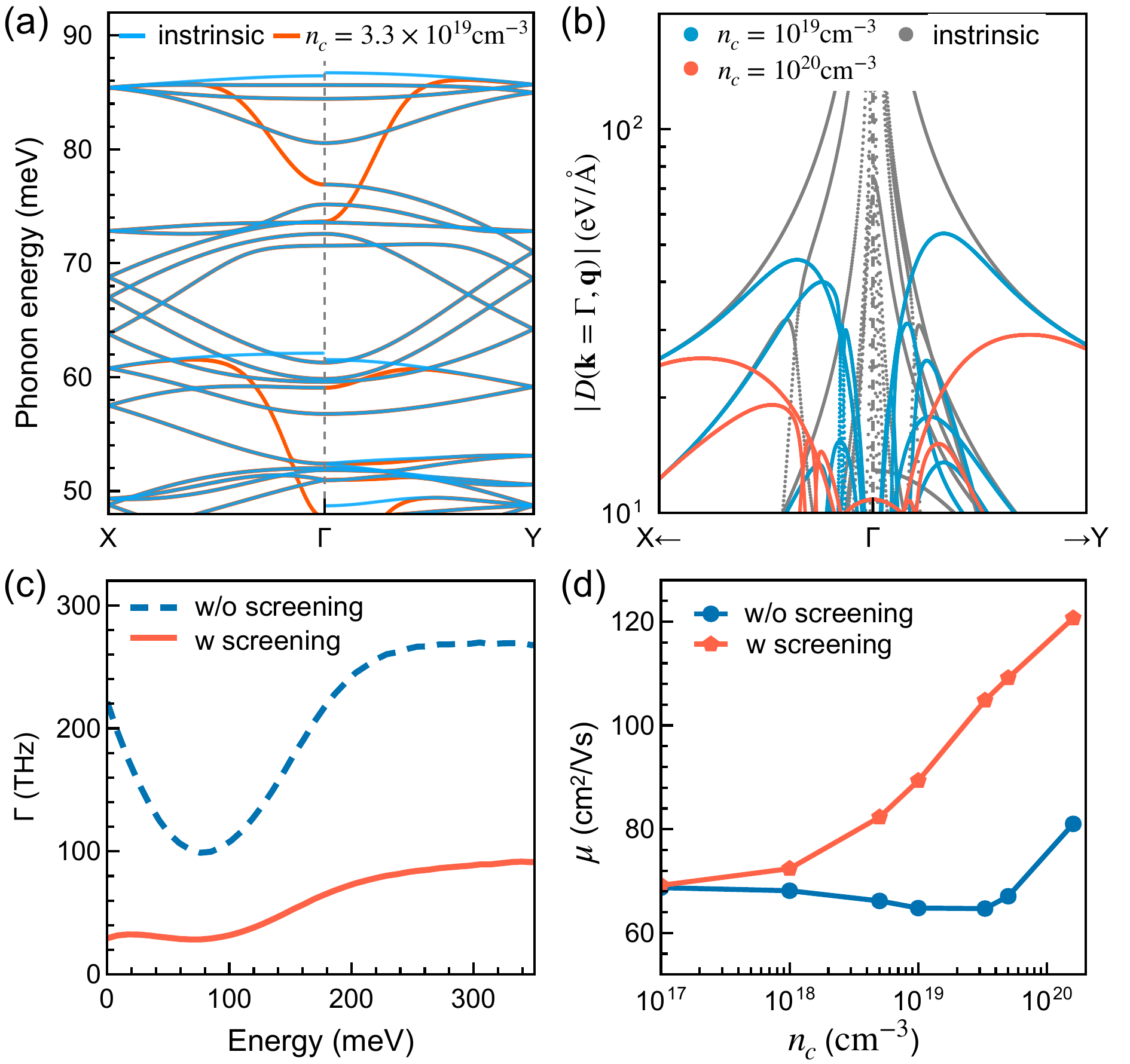}
\caption{Impacts of free-carrier screening. (a) Phonon dispersion in doped CaSnO$_3$ computed with free-carrier screening, and compared to the intrinsic system. 
(b) the e-ph deformation potential of the highest LO phonon branch computed at different doping levels. (c) the energy-dependent electron scattering rate at 300~K and $n_c \sim 10^{19} \text{cm}^{-3}$, and (d) the phonon-limited mobility as a function of carrier concentration with temperature fixed at 300~K, computed with (orange) and without (blue) free-carrier screening.}
\label{Fig:fig2}
\end{figure}

\label{sec:f1}
%
We computed the electronic band structure of CaSnO$_3$ using both the PBEsol functional and the more accurate HSE06 hybrid functional\cite{Krukau2006}. 
While the PBEsol yields a band gap of $\sim$2.5~eV, HSE06 predicts a band gap of about 4.1~eV, in closer agreement with the experimental value of $\sim$4.6~eV\cite{liu2023doping}. 
Nevertheless, the two functionals produce nearly identical conduction-band dispersions and electron effective masses near the conduction-band minimum, as illustrated in Fig.~\ref{Fig:fig1}(b).  For example, the effective mass along $\Gamma$-$Y$ is about 0.47~$m_e$ from PBEsol, while the value from HSE06 is about 0.46~$m_e$. 
This excellent agreement validates our use of PBEsol for subsequent transport calculations, where accurate effective masses are critical.

%
%
\indent
We first focus on e-ph interactions and electron transport in the intrinsic or lightly doped regime. 
The phonon dispersion of  CaSnO$_3$, shown in Fig.~\ref{Fig:fig1}(c), reveals a complex spectrum characteristic of the distorted perovskite structure containing 20 atoms per unit cell. 
The orthorhombic distortion from the ideal cubic perovskite structure results in Brillouin zone folding, causing each original phonon mode to split into four branches. 
The color scale indicates the mode-resolved e-ph coupling strength, $|g_{\nu}(\bm{q})|$, highlighting the dominant role of LO phonon modes. 
The LO modes at $\sim$87 meV and $\sim$62 meV exhibit the strongest coupling with electrons through long-range Fröhlich-type interactions, consistent with behavior observed in other perovskite oxides such as SrTiO$_3$~\cite{ZhouSTO} and SrSnO$_3$~\cite{truttmann_combined_2021}. We note that the long-range LO phonon coupling is insensitive to local distortions. \\
\indent
Figure~\ref{Fig:fig1}(d) presents the calculated phonon-limited electron mobility in lightly doped CaSnO$_3$, where free-carrier screening effects are negligible. The temperature dependence follows an approximate $T^{-2}$ trend, reaching $\sim$68~$\cvs$ at room temperature. 
This value is lower than that of orthorhombic SrSnO$_3$ ($\sim$76~$\cvs$)~\cite{truttmann_combined_2021}, which can be attributed to the larger electron effective mass in CaSnO$_3$ resulting from its enhanced octahedral distortions.
To quantify the relative importance of different scattering mechanisms, the inset of Fig.~\ref{Fig:fig1}(d) compares the total e-ph scattering rate with the contribution from long-range electron-LO phonon interactions. The near-complete overlap demonstrates the dominance of LO phonon scattering in intrinsic or lightly doped CaSnO$_3$, providing the physical basis for understanding how screening of these long-range interactions will impacts electron mobility in heavily doped samples.

\begin{figure}[!tbp]
    \centering
    \includegraphics[width=\columnwidth]{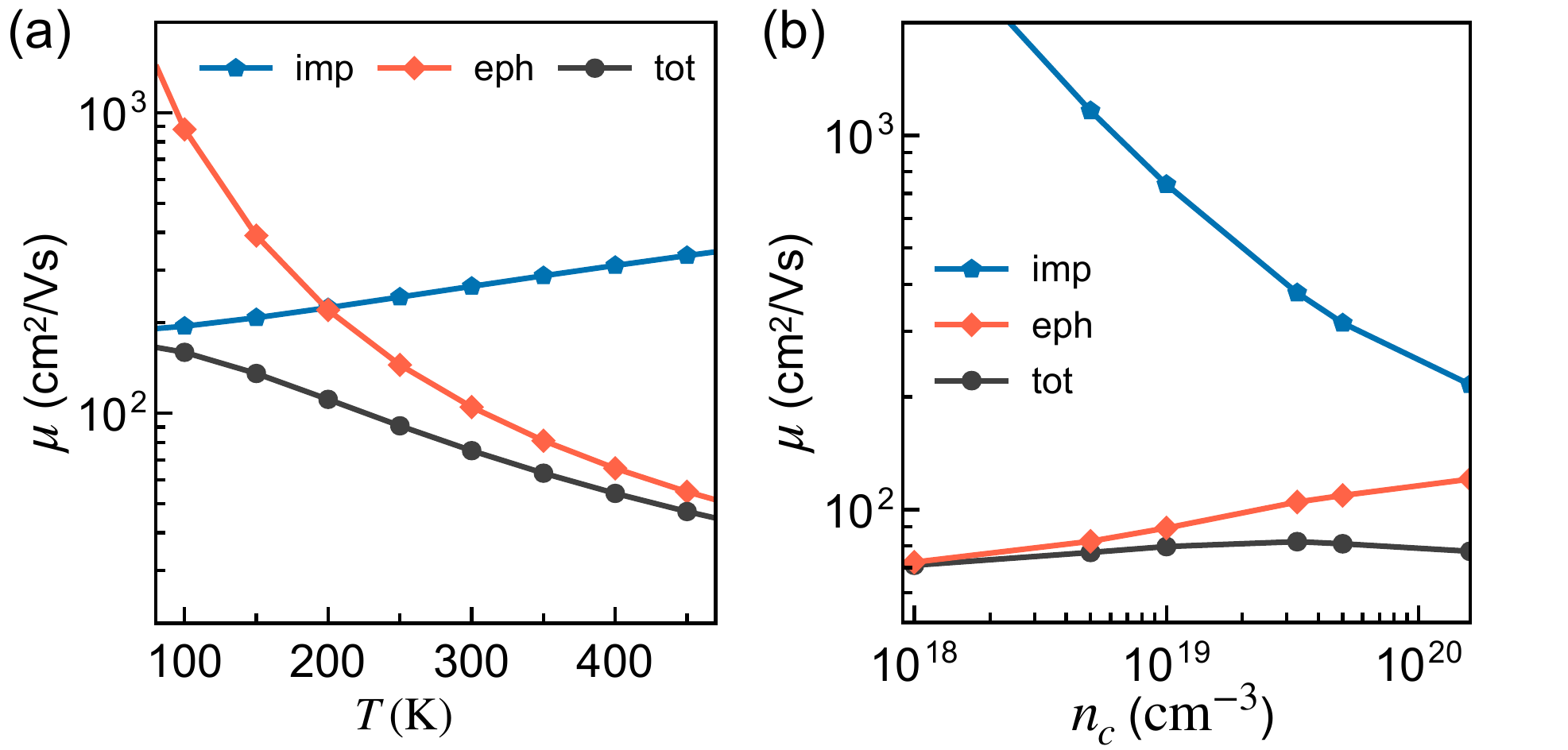}
    \caption{Electron mobility limited by ionized impurity scattering (blue), e-ph scattering (orange), and their combined effect (black): 
(a) as a function of temperature at $n_c \sim 3.3 \times 10^{19} \text{cm}^{-3}$, and (b) as a function of carrier concentration at 300~K.}
\label{Fig:fig3}
\end{figure}

We next examine how free-carrier screening modifies e-ph interactions and transport at elevated doping levels. 
Figure~\ref{Fig:fig2} systematically compares phonon dispersion, e-ph coupling, scattering rates, and electron mobility computed with and without screening effects.
The phonon dispersion changes dramatically with doping, as shown in Fig.~\ref{Fig:fig2}(a) for a representative carrier concentration of $3.3 \times 10^{19}~\text{cm}^{-3}$. 
In undoped CaSnO$_3$, the two highest LO branches exhibit large LO-TO splitting due to long-range macroscopic electric fields. With free carriers present, screening effectively suppresses these fields, causing the LO modes to soften and converge with TO modes at $\Gamma$ point, eliminating the LO-TO splitting while leaving other phonon branches largely unchanged.\\
\indent
This screening profoundly affects the e-ph coupling strength for the polar LO modes. To quantify the e-ph coupling strength at various doping concentrations, 
we compute the gauge-invariant e-ph deformation potential, $D_{\nu}(\bm{k},\bm{q})$, defined as:
\begin{align}
    D_{\nu}(\bm{k},\bm{q})=\sqrt{2\omega_{\nu\bm{q}}M_{\mathrm{tot}}} \left|g_{\nu}(\bm{k},\bm{q})\right| / \hbar, \label{eq:dfp}
\end{align}
where $\omega_{\nu\bm{q}}$ is the phonon frequency, $M_{\mathrm{tot}}$ is the total mass of the unit cell, and $|g_{\nu}(\bm{k},\bm{q})|$ is the absolute value of the e-ph matrix element for the lowest conduction band. 
Figure~\ref{Fig:fig2}(b) presents the e-ph deformation potential of the highest LO mode, which diverges as $1/q$ in the intrinsic regime---characteristic of long-range Fröhlich interactions. 
At $n_c\sim 10^{19} \text{cm}^{-3}$, this divergence is suppressed, with the coupling strength decreasing sharply and approaching zero as $q \rightarrow 0$. 
This behavior defines a characteristic screening wavevector $q_{\text{scr}}$ below which screening becomes effective. 
Higher carrier concentrations yield larger $q_{\text{scr}}$ and shorter screening lengths [see Fig.~\ref{Fig:fig2}(b)]. 
The suppressed polar coupling directly reduces e-ph scattering rates, as shown in Fig.~\ref{Fig:fig2}(c). 
The energy-dependent scattering rates decrease significantly when screening is included, particularly at energies relevant for transport, 
which stems from the screening of the long-range coupling between electron and LO phonons.
This reduction drives the mobility enhancement in the doped regime. \\

\indent
Figure~\ref{Fig:fig2}(d) compares the phonon-limited electron mobility at room temperature as a function of carrier concentration computed with and without screening.
At low carrier concentration of $\sim10^{17}~\text{cm}^{-3}$, screening effects are negligible and both calculations yield identical results. 
As doping increases, the difference becomes pronounced. Without screening, mobility initially decreases due to increased scattering phase space, then rises modestly beyond $\sim 3.3\times 10^{19}~\text{cm}^{-3}$ as the Fermi level shifts deeper into the conduction band --- with a mobility of $\sim64~\cvs$ at $3.3\times 10^{19}~\text{cm}^{-3}$ and $\sim80~\cvs$ at $1.6\times 10^{20}~\text{cm}^{-3}$. 
However, when screening effects are included, mobility increases steadily with doping, reaching $\sim105~\cvs$ at $3.3\times 10^{19}~\text{cm}^{-3}$ and $\sim120~\cvs$ at $1.6\times10^{20}~\text{cm}^{-3}$---nearly double the unscreened value. 
This demonstrates that free-carrier screening is essential for accurately predicting transport in heavily doped polar semiconductors.

To complete our analysis of electron transport in doped CaSnO$_3$, we assess the role of ionized impurity scattering at high carrier concentrations. 
The impurity-limited mobility is calculated using an effective mass model that incorporates an energy-dependent transport relaxation time, 
details of the computational method are provided in supplementary material.
The total mobility is obtained using Matthiessen's rule: $\mu^{-1}_{\text{tot}} = \mu^{-1}_{\text{eph}} + \mu^{-1}_{\text{imp}}$, where $\mu_{\text{eph}}$ and $\mu_{\text{imp}}$ represent phonon- and impurity-limited mobilities, respectively.
Figure~\ref{Fig:fig3} illustrates the interplay between these scattering mechanisms. At a representative doping level, the ionized impurity scattering dominates below ~200 K, while e-ph scattering controls mobility above room temperature [see Fig.~\ref{Fig:fig3}(a)]. 
The carrier concentration dependence at 300~K, shown in Fig.~\ref{Fig:fig3}(b), reveals that e-ph scattering remains dominant across most doping levels, but impurity scattering becomes increasingly important beyond $\sim 3.3\times 10^{19}~\text{cm}^{-3}$, partially offsetting the screening-induced mobility enhancement. 

\begin{figure}
    \centering
    \includegraphics[width=\columnwidth]{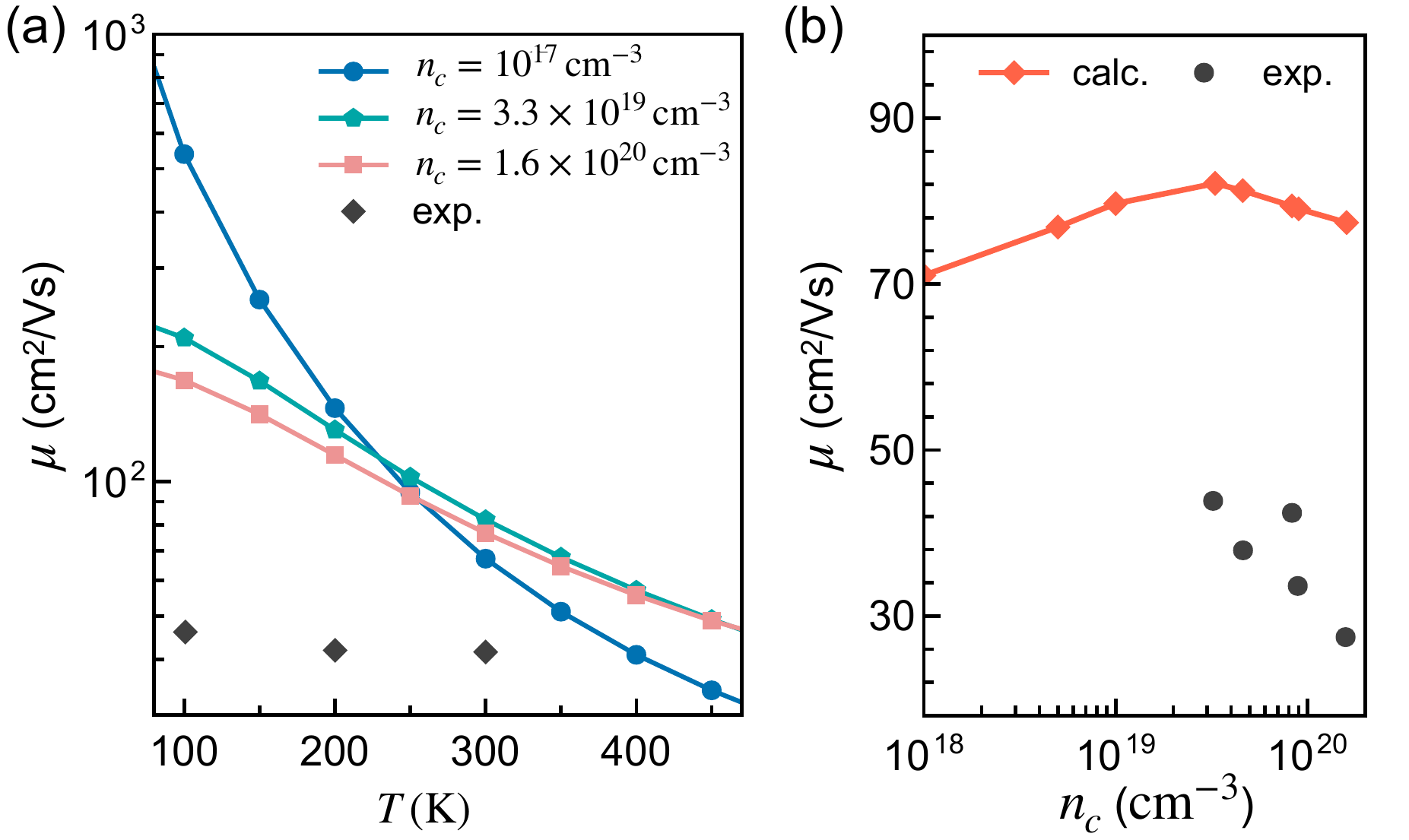}
    \caption{
Computed electron mobility including both e-ph scattering with free-carrier screening and ionized impurity scattering, compared to experimental data: (a) temperature-dependent mobility at various carrier concentrations, and (b) $n_c$-dependent mobility at room temperature. Experimental values are taken from Ref.~\onlinecite{liu2023doping}.}
\label{Fig:fig4}
\end{figure}

Finally, we compare our calculated mobilities with recent experimental data reporting the highest values to date, as shown in Fig.~\ref{Fig:fig4}. The combined effects of screening and impurity scattering produce distinct behavior in heavily doped samples: lower mobility below ~250 K but higher mobility at elevated temperatures compared to intrinsic systems. 
The calculated mobility shown in Fig.~\ref{Fig:fig4}(a) exhibits weaker temperature dependence (approximately $T^{-1}$) under heavy doping, compared to the intrinsic $\sim T^{-2}$ dependence, while experimental data show minimal temperature variation with only slight decrease at higher temperatures. 
The room-temperature mobility versus carrier concentration [Fig.~\ref{Fig:fig4}(b)] further reveals this discrepancy. 
Our predicted mobility initially rise due to screening effects, peak at $\sim83~\cvs$ near $n_c = 3.3\times 10^{19} \text{cm}^{-3}$, then decline due to impurity scattering.
However, experimental values remain substantially lower---only $42~\cvs$ at the concentration of $3.3\times10^{19} \text{cm}^{-3}$---and decrease more rapidly with increased doping. At $n_c = 1.6 \times 10^{20} \text{cm}^{-3}$, the gap widens further: we predict $\sim73~\cvs$ versus the experimental $25~\cvs$\cite{liu2023doping}. We note that our prediction does not take into account potential higher-order effects, such as anharmonic effects and carrier localization, which may also influence the mobility.\\
\indent
The minimal temperature variation and stronger doping dependence of experimental mobilities suggest dominance by extrinsic scattering from dislocation, grain boundaries, and various defects including oxygen vacancies, rather than intrinsic phonon scattering. 
Our results suggest considerable untapped potential: room-temperature mobilities could reach twice current experimental values through improved material quality. 
Moreover, advanced doping strategies like modulation doping~\cite{Dingle1978} could potentially achieve the phonon-limited mobility of $\sim120~\cvs$ by spatially separating carriers from ionized impurities.

\label{summary}
In summary, we have performed first-principles calculations of electron mobility in doped CaSnO$_3$ using state-of-the-art methods that explicitly account for free-carrier screening. Our results reveal that transport is dominated by long-range polar optical phonon scattering, which becomes significantly suppressed at high carrier concentrations due to screening. 
This screening effect drives substantial mobility enhancement, with phonon-limited values reaching $\sim$120~$\cvs$ at room temperature.
While ionized impurity scattering emerges as a competing mechanism at high doping, the net mobility at room-temperature still increases significantly---from $\sim$68~$\cvs$ in intrinsic material to $\sim$83~$\cvs$ at optimal doping when both mechanisms are considered, about twice the highest experimental values. 
This large gap indicates that current samples are limited by extrinsic defects rather than intrinsic scattering. 
Our findings establish the fundamental transport limits of CaSnO$_3$ and demonstrate that improved material quality could unlock exceptional conductivity while maintaining its ultrawide bandgap, positioning it as a compelling candidate for transparent and high-power electronics.

\section*{Supplementary Material}
See the supplementary material for more details of the convergence tests and a more comprehensive description of our computational methods for transport calculations including free-carrier screening and ionized impurity scattering.

\begin{acknowledgments} 
The authors acknowledge support from the National Natural Science Foundation of China (Grant Nos.~12104039 and~42230311), the Beijing Natural Science Foundation (Grant No.~Z210006), and the National Key R\&D Program of China (Grant No.~2022YFA1403400). 
\end{acknowledgments}

\section*{AUTHOR DECLARATIONS}
\section*{Conflict of Interest}
The authors have no conflicts to disclose.
\section*{DATA AVAILABILITY}
The data that support the findings of this study are available from the corresponding author upon reasonable request.

\bibliographystyle{apsrev4-2}
\bibliography{cso_ref}

\end{document}